\documentclass[pra,groupedaddress,showpacs,floatfix]{revtex4}

\begin{document}

\title{A semi-quantum version of the game of Life}
\author{A. P. Flitney}
\email{aflitney@eleceng.adelaide.edu.au}
\author{D. Abbott}
\email{dabbott@eleceng.adelaide.edu.au}
\affiliation{Centre for Biomedical Engineering (CBME)
and Department of Electrical and Electronic Engineering, \\
The University of Adelaide, SA 5005, Australia}
\date{\today}

\begin{abstract}
A version of John Conway's game of Life is presented
where the normal binary values of the cells are replaced by oscillators which
can represent a superposition of states.
The original game of Life is reproduced in the classical limit,
but in general additional properties not seen in the original game are 
present that display some of the effects of a quantum mechanical Life.
In particular, interference effects are seen.
\end{abstract}


\maketitle

\section{Introduction}
John Conway's game of Life~\cite{gar71} is a well known two dimensional
cellular automaton where cells are arranged in a square grid and have 
binary values generally known as dead or alive.
The status of the cells change in a discrete fashion each ``generation'' 
depending upon the number of neighboring cells that are alive, the 
general idea being that a cell dies if there is either overcrowding or 
isolation. There are many different rules that can be applied for 
birth or survival of a cell and a number of these give rise to 
interesting properties such as still lives (stable patterns), 
oscillators (patterns that periodically repeat),
spaceships or gliders (fixed shapes that move across the Life universe),
glider guns, and so on~\cite{gar71,ber82,gar83}.
Conway's original rules are one of the few that are balanced between 
survival and extinction of the life ``organisms.''
In this version a dead (or empty) cell becomes alive if it has exactly 
three living neighbors,
while an alive cell survives if and only if it has two or three living 
neighbors.
Much literature on the game of Life and its implications exists.
For recent discussion on the possibilities of this
and other cellular automata the interested reader is referred to~\cite{wol02}.

The recent interest in quantum games
\cite{mey99,eis99,mar00,ben00a,ben00b,john01,du01,iqbal01,iqbal02,flitney02a,flitney02b}
suggests the possibility of applying the idea of superposition of states in
quantum mechanics to the game of Life.
Unfortunately Conway's Life is irreversible while,
in the absence of a measurement,
quantum mechanics is reversible.
In particular, operators that represent measurable quantities
must be unitary.
A full quantum Life would be problematic given the known difficulties
of quantum cellular automata~\cite{mey96}.
Interesting behavior can still be obtained in a semi-quantum mechanical Life
by representing the cells by classical sine-wave oscillators
with a period equal to one generation,
an amplitude between zero and one,
and a variable phase.
The amplitude of the oscillation represents the coefficient of the alive state
so that the square of the amplitude gives the probability
of finding the cell in the alive state when a measurement of the ``health''
of the cell is taken.
If the initial state of the system contains at least one
cell that is in a superposition of eigenstates
the neighboring cells will be influenced according to the coefficients 
of the respective eigenstates,
propagating the superposition to the surrounding region.

If the coefficients of the superpositions are restricted to positive 
real numbers we simply obtain a probabilistic Life universe (i.e. a 
classical mixture of states) where we do not expect to see qualitatively new
phenomena.
By allowing the coefficients to be complex,
that is, by allowing phase differences between the oscillators,
qualitatively new phenomena,
for example interference effects, may arise.

\section{Classical Mixture of States}
To represent the state of a cell we introduce the following notation:
\begin{equation}
|\psi\rangle = a |\mbox{alive}\rangle + b |\mbox{dead}\rangle \;,
\end{equation}
subject to the normalization condition
\begin{equation}
|a|^2 + |b|^2 = 1 \;.
\end{equation}
$|a|^2$ and $|b|^2$ represent the probabilities of measuring the cell as alive
or dead, respectively.
If the values of $a$ and $b$ are restricted to non-negative real numbers the
cell is a classical mixture of the alive and dead states.
In our model $|a|$ is the amplitude of the oscillator.
Restricting $a$ to non-negative real numbers corresponds to the oscillators all
being in phase.

The birth, death and survival operators have the following effects
\begin{eqnarray}
B |\psi\rangle &=& |\mbox{alive}\rangle \\ \nonumber
D |\psi\rangle &=& |\mbox{dead}\rangle \\ \nonumber
S |\psi\rangle &=& |\psi\rangle \;\;.
\end{eqnarray}
A cell can be represented by the vector
\begin{displaymath}
\left( \begin{array}{c} a \\ b \end{array} \right) \;.
\end{displaymath}
The $B$ and $D$ operators are not unitary.
Indeed they can be represented in matrix form as
\begin{eqnarray}
B &=& \left( \begin{array}{cc} 1 & 1 \\ 0 & 0 \end{array} \right) \;, \\
D &=& \left( \begin{array}{cc} 0 & 0 \\ 1 & 1 \end{array} \right) \;.
\end{eqnarray}

A new generation is obtained by determining the number of living neighbors
each cell has
and then applying the appropriate operator to that cell.
The number of living neighbors in our model is the amplitude of the
superposition of the oscillators
representing the surrounding eight cells.
This process is carried out on all cells effectively simultaneously.
When the cells are permitted to take a superposition of states,
the number of living neighbors need not be an integer.
Thus a mixture of the $B$, $D$ and $S$ operators may need to be applied.
For consistency with standard Life the following conditions will be imposed
upon the operators that produce the next generation:
\begin{itemize}
\item
If there are an integer number of living neighbors the operator applied must
be the same as that in standard Life.
\item
The operator that is applied to a cell must continuously change
from one of the basic forms to another
as the sum of the $a$ coefficients from the neighboring cells changes
from one integer to another.
\item
The operators can only depend upon this sum
and not on the individual coefficients.
\end{itemize}
If the sum of the $a$ coefficients of the surrounding eight cells is
\begin{equation}
A = \sum_{i=1}^{8} \, a_i
\end{equation}
then the following set of operators,
depending upon the value of $A$,
is the simplest that has the required properties
\begin{eqnarray}
\label{eqn-gen}
0 \le A \le 1; \; & G_0 =& D \;, \\ \nonumber
1 < A \le 2; \; & G_1 =& (\sqrt{2} +1) (2 - A) D \:+\: (A - 1) S \;, \\ \nonumber
2 < A \le 3; \; & G_2 =& (\sqrt{2} +1) (3 - A) S \:+\: (A - 2) B\;,  \\ \nonumber
3 < A < 4; \; & G_3 =& (\sqrt{2} +1) (4 - A) B \:+\: (A - 3) D \;, \\ \nonumber
A \ge 4; \; & G_4 =& D \;.
\end{eqnarray}
For integer values of $A$, the $G$ operators are the same as the basic operators
of standard Life, as required.
For non-integer values in the range $(1,4)$,
the operators are a linear combination of the standard operators.
The factors of $\sqrt{2} + 1$ have been inserted to give more appropriate
behavior in the middle of each range.
For example, consider the case where $A = 3 + 1/\sqrt{2}$,
a value that may represent three neighboring cells that are alive
and one the has a probability of one half of being alive.
The operator in this case is
\begin{equation}
G = \frac{1}{\sqrt{2}} \, B + \frac{1}{\sqrt{2}} \, D \;.
\end{equation}
Applying this to either a cell in the alive,
$\left( \begin{array}{cc} 1 \\ 0 \end{array} \right)$
or dead,
$\left( \begin{array}{cc} 0 \\ 1 \end{array} \right)$
states will produce the state
\begin{equation}
|\psi\rangle = \frac{1}{\sqrt{2}} \, |\mbox{alive}\rangle + \frac{1}{\sqrt{2}}
\, |\mbox{dead}\rangle
\end{equation}
which represents a cell with a 50\% probability of being alive.
That is, $G$ is an equal combination of the birth and death
operators, as we might have expected given the possibility that $A$ 
represents an equal probability of three or four living neighbors.
Of course the same value of $A$ may have been obtained by other 
combinations of neighbors that do not lie half way between three and 
four living neighbors,
but one of our requirements is that the operators can only depend on the sum of
the $a$ coefficients of the neighboring cells
and not on how the sum was obtained.

The new state of a cell is obtained by calculating $A$,
applying the matrix $G$
corresponding to the appropriate operator:
\begin{equation}
\left( \begin{array}{c}
	a' \\
	b'
\end{array} \right)		= G \left( \begin{array}{c}
				a \\
				b
			\end{array} \right) \;,
\end{equation}
and then normalizing the resulting state so that $|a'|^2 +|b'|^2 = 1$.
It is this process of normalization that means that multiplying the 
matrix by a constant has no effect.
Hence, for example, $G_2$ for $A=3$ has the same effect as $G_3$ in 
the limit as $A \rightarrow 3$, despite differing by the constant factor 
$(\sqrt{2} + 1)$.

\section{Semi-quantum Life}
To get qualitatively different behavior from classical Life we need to 
introduce a phase associated with the coefficients,
that is, a phase difference between the oscillators.
We require the following features from this version of Life:
\begin{itemize}
\item
It must smoothly approach the classical mixture of states
as all the phases are taken to zero.
\item
Interference, that is the partial or complete cancellation between
cells of different phases, must be possible.
\item
The overall phase of the Universe must not be measurable.
That is, multiplying all cells by $e^{i \phi}$ for some real $\phi$ will have
no measurable consequences.
\item
The symmetry between $(B,\: |\mbox{alive}\rangle)$ and $(D,\:
|\mbox{dead}\rangle)$
that is a feature of the original game of Life
should be retained.
That is, if the state of all cells is reversed
($|\mbox{alive}\rangle \longleftrightarrow |\mbox{dead}\rangle$)
and the operation of the $B$ and $D$ operators is reversed
the system will behave in the same manner.
%While this symmetry is not essential
%our aim is to preserve as many of the features of the original game as 
%possible.
\end{itemize}
In order to incorporate complex coefficients while keeping the above 
properties, the basic operators are modified in the following way:
\begin{eqnarray}
B |\mbox{dead}\rangle &=& e^{i \phi} |\mbox{alive}\rangle \;, \\ \nonumber
B |\mbox{alive}\rangle &=& |\mbox{alive}\rangle \;, \\ \nonumber
D |\mbox{alive}\rangle &=& e^{i \phi} |\mbox{dead}\rangle \;, \\ \nonumber
D |\mbox{dead}\rangle &=& |\mbox{dead}\rangle \;, \\ \nonumber
S |\psi\rangle &=& |\psi\rangle \;,
\end{eqnarray}
where the superposition
of the surrounding oscillators is
\begin{equation}
\alpha = \sum_{i=1}^{8} \, a_i = A e^{i \phi} \;,
\end{equation}
$A$ and $\phi$ being real positive numbers.
That is, the birth and death operators are modified
so that the new alive or dead state
has the phase of the sum of the surrounding cells.
The operation of the $B$ and $D$ operators on the state
$\left( \begin{array}{cc} a \\ b \end{array} \right)$
can be written as
\begin{eqnarray}
B \left( \begin{array}{cc} a \\ b \end{array} \right)
	&=& \left( \begin{array}{cc} a + |b| e^{i \phi} \\ 0 \end{array}
\right) \;, \\ \nonumber
D \left( \begin{array}{cc} a \\ b \end{array} \right)
	&=& \left( \begin{array}{cc} 0 \\ |a| e^{i \phi} + b \end{array}
\right) \;,
\end{eqnarray}
with $S$ leaving the cell unchanged.
The presence of the modulus signs means that
the $B$ and $D$ operators are not linear.
The modulus of the sum of the neighboring cells, $A$, determines which 
operators apply, in the same way as before
(see Eqn. (\ref{eqn-gen})).
The addition of the phase factors for the cells allows for interference 
effects since the coefficients of alive cells may not always reinforce 
in taking the sum, $\alpha = \sum a_i$.
A cell with $a = -1$ still has a unit probability of being measured in 
the alive state but its effect on the sum will cancel that of a cell with $a =
1$.
It will be noted that the phase of the dead state has no effects.
It is retained in order to maintain the alive $\longleftrightarrow$ dead
symmetry.

A useful notation to represent semi-quantum Life is to use an arrow whose 
length represents the amplitude of the $a$ coefficient and whose angle 
with the horizontal is a measure of the phase of $a$.
That is, the arrow represents the phasor of the oscillator at the
beginning of that generation.
For example
\begin{eqnarray}
\longrightarrow &=&	\left( \begin{array}{c}
			1 \\
			0
		\end{array} \right) \;, \\ \nonumber
\nearrow &=&	e^{i \pi/4} \left( \begin{array}{c}
				   1/\sqrt{2} \\
				   1/\sqrt{2}
			   \end{array} \right)
		      = \left( \begin{array}{c}
			      (1+i)/\sqrt{2} \\
			      (1+i)/\sqrt{2}
		      \end{array} \right) \;, \\ \nonumber
\end{eqnarray}
etc.
Then $\alpha$ is the vector sum of the arrows.
This notation includes no information about the $b$ coefficient.
The magnitude of this coefficient can be determined from $a$ and the 
normalization condition.
As noted previously, the phase of the $b$ coefficient has no effect on the
future progression of the game so it is not necessary to represent this.

\section{Results and Conclusions}
\begin{figure}
\setlength{\unitlength}{0.8pt}
\begin{picture}(280,150)(-50,-20)

	% (a) light-speed wick with a block at one end

	\put(0,100)
	{ \begin{picture}(220,50)(0,0)
		\put(-50,0){(a)}
		% initital block
		\put(0,0){\vector(1,0){20}}
		\put(25,0){\vector(1,0){20}}
		\put(0,25){\vector(1,0){20}}
		\put(25,25){\vector(1,0){20}}

		% wick
		\multiput(70,0)(50,0){3}{\vector(-1,0){20}}
		\multiput(75,0)(50,0){3}{\vector(1,0){20}}
		\put(200,0){\ldots}
	\end{picture} }
	
	% (b) double thickness light-speed wick

	\put(0,0)
	{ \begin{picture}(220,50)(0,0)
		\put(-50,10){(b)}
		% stabilise the left hand end
		\put(10,0){\vector(0,1){20}}
		\put(10,25){\vector(0,1){20}}

		% double thickness wick
		\multiput(70,10)(50,0){3}{\vector(-1,0){20}}
		\multiput(70,35)(50,0){3}{\vector(-1,0){20}}
		\multiput(25,10)(50,0){4}{\vector(1,0){20}}
		\multiput(25,35)(50,0){4}{\vector(1,0){20}}
		\put(200,10){\ldots}
		\put(200,35){\ldots}
	\end{picture} }
\end{picture}

\caption{(a) A wick
(an extended structure that dies, or ``burns'', at a constant rate)
that burns at the speed of light
(one cell per generation).
The block on the left-hand end stabilises that end;
a block on both ends would give a stable line;
the absence of the block would give a wick that burns from both ends.
(b) Another example of a light-speed wick in semi-quantum Life
showing one method of stabilising the left-hand end.}
\label{fig-wick}
\end{figure}
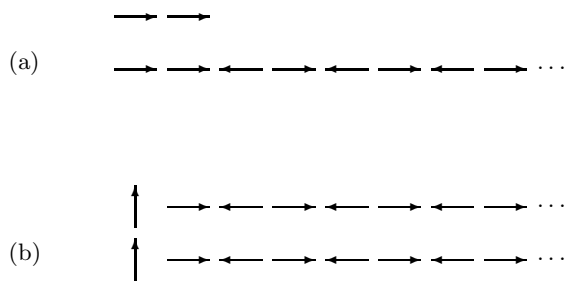

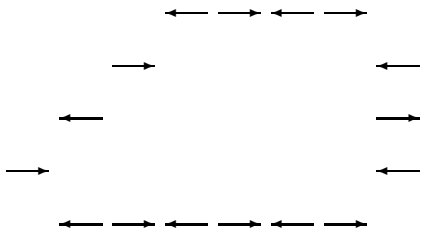
\begin{figure}
\setlength{\unitlength}{0.8pt}
\begin{picture}(220,120)(-25,-20)
	% stable loop of string
	\multiput(20,0)(50,0){3}{\vector(-1,0){20}}
	\multiput(25,0)(50,0){3}{\vector(1,0){20}}
	\multiput(70,100)(50,0){2}{\vector(-1,0){20}}
	\multiput(75,100)(50,0){2}{\vector(1,0){20}}
	\multiput(170,25)(0,50){2}{\vector(-1,0){20}}
	\put(150,50){\vector(1,0){20}}
	\multiput(-25,25)(50,50){2}{\vector(1,0){20}}
	\put(20,50){\vector(-1,0){20}}
\end{picture}

\caption{An example of a stable loop made from cells of alternatring phase.
Above a certain minimium, such structures can be made of arbitary size and
shape.}
\label{fig-string}
\end{figure}

\begin{figure}
\setlength{\unitlength}{0.8pt}
\begin{picture}(220,100)(-25,-20)
	% stable boundary by adjusting phases
	\multiput(0,30)(25,0){7}{\vector(1,0){20}}
	\multiput(40,0)(100,0){2}{\vector(-2,3){10}}
	\put(90,15){\vector(-2,-3){10}}
	\multiput(30,45)(100,0){2}{\vector(2,3){10}}
	\put(80,60){\vector(2,-3){10}}
	\put(175,30){\ldots}
	\put(-25,30){\ldots}
\end{picture}

\caption{A boundary utilising appropriate phase differences to
produce stability.
The upper cells are out of phase by $\pm \pi/3$
and the lower by $\pm 2 \pi/3$ with the central line.}
\label{fig-boundary}
\end{figure}
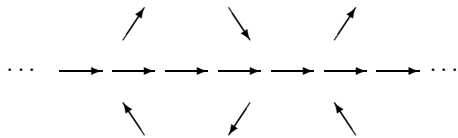

The above rules have been implemented in {\em Mathematica} \cite{math}.
Some simple results are indicated in the attached figures.
Figure \ref{fig-block} shows an elementary example of interference in a block
still life with one cell out of phase with its neighbors.
Figures \ref{fig-wick} and \ref{fig-string} show the sort of structures,
a wick and a loop that can be obtained from a line of cells of alternating phase,
that is of units of $\longrightarrow \longleftarrow$'s.
The stability of the line of $\longrightarrow \longleftarrow$'s
results from the fact that while each cell in the line has exactly
two living neighbors,
the cells above or below this line have a net of zero
(or one at a corner) living neighbors,
due to the canceling effect of the opposite phases.
No new births around the line will occur
unlike the case where all the cells are in phase.
The equivalent structure in Conway's Life is a diagonal line of live cells,
but in our case we are able to make the loop any shape.
Figure \ref{fig-boundary}
shows a stable boundary that results from the appropriate adjustment
of the phase differences between the cells.
The angles have been chosen so that each cell in the line
has between two and three living neighbors,
while the empty cells above and below the line
have either two or four living neighbors and so remain life-less.
Wicks and boundaries are known in standard Life
but generally require more complex structures.

To summarize,
John Conway's game of Life is a two dimensional cellular automaton
where the new state of a cell is determined by the sum of the neighboring
states that are in one particular state generally referred to as ``alive''.
In semi-quantum Life cells may be in a superposition of the
alive and dead states
with the coefficient of the alive state being represented by an oscillator.
The equivalent of evaluating the number of living
neighbors of a cell is to take the superposition of the oscillators of the
surrounding states.
The amplitude of this superposition
will determine which operator(s) to apply
to the central cell to determine its new state,
while the phase gives the phase of any new state produced.
Such a system is able to reproduce some of the aspects of quantum
mechanics such as interference.

Obviously this paper just touches on some of the results
that can be obtained with this new scheme.
More detailed possibilities will not emerge until a systematic search of the
semi-quantum Life possibilities is undertaken
but it can be seen that some new effects and structures occur
and that some of the known effects in Conway's Life
can occur in a simpler manner.

\begin{figure}[h]
\setlength{\unitlength}{0.8pt}
\begin{picture}(200,350)

	% block with one cell out of phase

	% (a)
	\put(0,250)
	{ \begin{picture}(220,100)(-50,0)
		\put(0,25)
		{ \begin{picture}(50,50)
			% initital block
			\put(0,0){\vector(1,0){20}}
			\put(45,0){\vector(-1,0){20}}
			\put(0,25){\vector(1,0){20}}
			\put(25,25){\vector(1,0){20}}
		\end{picture} }
		\put(0,0){initial}
		\put(-50,25){(a)}

		%first generation
		\put(80,25)
		{ \begin{picture}(50,50)
			\put(45,0){\vector(-1,0){20}}
			\multiput(10,0)(0,25){2}{\circle*{2}}
			\put(35,25){\circle*{2}}
		\end{picture} }
		\put(80,0){1st gen.}

		%second generation
		\put(160,0)
		{ \begin{picture}(50,50)
			\multiput(10,25)(0,25){2}{\circle*{2}}
			\multiput(35,25)(0,25){2}{\circle*{2}}
		\end{picture} }
		\put(160,0){2nd gen.}
	\end{picture} }

	% (b)
	\put(0,150)
	{ \begin{picture}(220,50)(-50,0)
		\put(0,0)
		{ \begin{picture}(50,50)
			\put(0,10){\vector(1,0){20}}
			\put(35,0){\vector(0,1){20}}
			\put(0,35){\vector(1,0){20}}
			\put(25,35){\vector(1,0){20}}
			\put(15,-25){(i)}
		\end{picture} }
		\put(-50,0){(b)}

		\put(100,0)
		{ \begin{picture}(50,50)
			\put(0,10){\vector(1,0){20}}
			\put(40,5){\vector(-2,3){9}}
			\put(0,35){\vector(1,0){20}}
			\put(25,35){\vector(1,0){20}}
			\put(45,10){${\scriptscriptstyle 2 \pi/3}$}
			\multiput(45,5)(6,0){3}{\line(1,0){3}}
		\put(15,-25){(ii)}
		\end{picture} }
	\end{picture} }

	% (c)
	\put(0,25)
	{ \begin{picture}(220,50)(-50,0)
		\put(-10,25)
		{ \begin{picture}(50,50)
			% initital block
			\put(0,0){\vector(1,0){20}}
			\put(42,-5){\vector(-1,1){14}}
			\put(0,25){\vector(1,0){20}}
			\put(25,25){\vector(1,0){20}}
			\put(45,0){${\scriptscriptstyle 3 \pi/4}$}
			\multiput(47,-5)(6,0){3}{\line(1,0){3}}
		\end{picture} }
		\put(-10,0){initial}
		\put(-50,25){(c)}

		%first generation
		\put(90,25)
		{ \begin{picture}(50,50)
			\put(0,0){\vector(1,0){18}}
			\put(42,-5){\vector(-1,1){14}}
			\put(0,25){\vector(1,0){18}}
			\put(25,25){\vector(1,0){18}}
		\end{picture} }
		\put(80,0){1st gen.}

		%second generation
		\put(170,25)
		{ \begin{picture}(50,50)
			\put(0,0){\vector(1,0){10}}
			\put(42,-5){\vector(-1,1){14}}
			\put(0,25){\vector(1,0){10}}
			\put(25,25){\vector(1,0){10}}
		\end{picture} }
		\put(170,0){2nd gen.}

		%third generation
%	\put(260,25)
%		{ \begin{picture}(50,50)
%			\put(0,0){\circle*{2}}
%			\put(35,-5){\vector(-1,1){14}}
%			\put(0,25){\circle*{2}}
%			\put(25,25){\circle*{2}}
%		\end{picture} }
%		\put(260,0){3rd gen.}
	\end{picture} }
\end{picture}

\caption{(a) A simple example of destructive interference
in semi-quantum Life:
a block with one cell out of phase by $\pi$ dies in two generations.
(b) Blocks where the phase difference of the fourth cell is insufficient
to cause complete destructive interference;
each cell maintains a net of at least two living neighbors
and so the patterns are stable.
In the second of these, the fourth cell is at a critical angle.
Any greater phase difference causes instability
resulting in eventual death as seen in (c),
which dies in the fourth generation.}
\label{fig-block}
\end{figure}
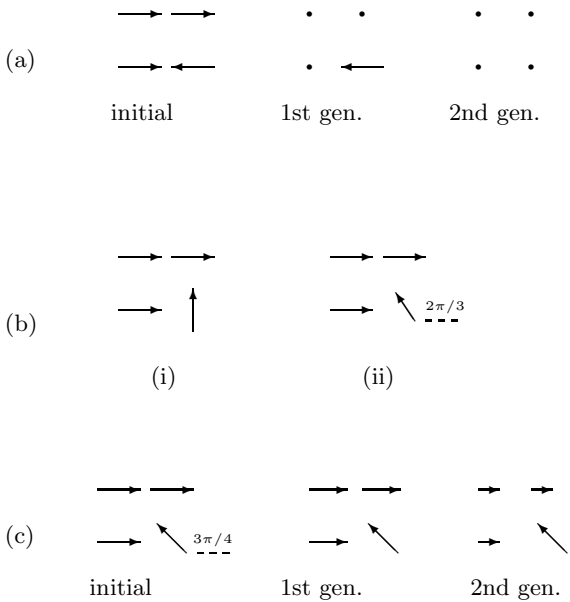

\end{document}